**Improved sequentially processed Cu(In,Ga)(S,Se)$_2$ by Ag alloying**

*Aubin JC. M. Prot\*, Michele Melchiorre, Tilly Schaaf, Ricardo G. Poeira, Hossam Elanzeery, Alberto Lomuscio, Souhaib Oueslati, Anastasia Zelenina, Thomas Dalibor, Gunnar Kusch, Yucheng Hu, Rachel A. Oliver, Susanne Siebentritt*

Aubin JC. M. Prot, Michele Melchiorre, Tilly Schaaf, Susanne Siebentritt
Laboratory for Photovoltaics, Physics and Materials Science Research Unit, University of Luxembourg, 41 rue du Brill, Belvaux L-4422, Luxembourg
E-mail: aubin.prot@uni.lu

Ricardo G. Poeira
Laboratory for Energy Materials, Physics and Materials Science Research Unit, University of Luxembourg, 41 rue du Brill, Belvaux L-4422, Luxembourg

Hossam Elanzeery, Alberto Lomuscio, Souhaib Oueslati, Anastasia Zelenina, Thomas Dalibor
AVANCIS GmbH, Otto-Hahn-Ring 6, 81739 München, Germany

Gunnar Kusch, Yucheng Hu, Rachel A. Oliver
Department of Materials Science and Metallurgy, University of Cambridge, 27 Charles Babbage Road, Cambridge CB3 0FS, UK

**Abstract**

Alloying small quantities of silver into Cu(In,Ga)Se$_2$ was shown to improve the efficiency for wide and low band gap solar cells. We study low band gap industrial Cu(In,Ga)(S,Se)$_2$ absorbers, substituting less than 10% of the copper with silver, using absolute photoluminescence and cathodoluminescence spectroscopy. Silver improves the grain size and promotes the interdiffusion of Ga and In across the depth of the absorber, resulting in a smoother band gap gradient. However, a certain lateral inhomogeneity is observed near the front and back sides. The non-radiative losses in the bare absorbers are reduced by up to 30 meV.

## 1. Introduction

Reaching a record power conversion efficiency (PCE) of 23.6% [1], chalcopyrite solar cells are one of the best thin film technologies for solar power harvesting. The efficiency record has been achieved with a Cu(In,Ga)Se$_2$ (CIGS) based solar cell, in which some of the Cu was replaced by silver (Ag). It is generally observed that Ag improves the optoelectronic quality of the CIGS solar cell. [2] Alloying with Ag decreases both the conduction ($E_C$) and valence ($E_V$) band energy. [3] $E_V$ decreases first slower than $E_C$ until a certain [Ag]/([Ag]+[Cu]) ratio (or AAC) is reached before the opposite happens. This results in a decrease of the band gap for low AAC and an increase for high AAC. [3] Depending on the [Ga]/([Ga]+[In]) ratio (or GGI), the change in band gap is more or less pronounced. According to Keller *et al.*, a small amount of Ag (AAC < 0.25) has a stronger effect (decrease of $E_g$) on a wider band gap material than on a lower band gap one, whose $E_g$ is barely affected at low AAC. [3]

Growing Ag-based solar cell has been done for nearly two decades already. In 2005, Nakada *et al.* achieved a 9.2% PCE for a 1.7 eV band gap Ag(In,Ga)Se$_2$ solar cell. [4] More recently, several studies reported on alloying Ag with wide band gap Cu(In,Ga)Se$_2$ (ranging from $E_g = 1.24$ eV to $E_g = 1.46$ eV) during co-evaporation [5–9], improving the performances mostly by increasing the $V_{OC}$. Improving such wide gap material would be interesting for multi-junction devices. At the same time, the effect of Ag-alloying is also actively investigated in lower band gap Cu(In,Ga)Se$_2$ (ranging from $E_g = 1.02$ eV to $E_g = 1.18$ eV). [10–14] Leading to a lower melting temperature than the pure Cu compounds [15], alloying with Ag offers the possibility to process the absorbers at lower temperature, which represents a great advantage from an industrial point of view. Alternatively, growing at similar temperatures as the Ag-free compound, grain size can be increased [15] and crystal defects may be reduced as simulated in Zhang *et al.*. [16] The presence of Ag in the crystal lattice has the following reported effects. An improved morphology is widely observed, translated into an increase of the grain size [17], a smoother surface [14] and, in low band gap absorbers, a passivation of deep defects [18]. In addition, a reduction of the doping – hole density – as well as an improved current collection are systematically reported and attributed to a widening of the space chare region. [12] The optimum amount of Ag necessary to obtain improved device performance depends strongly on the composition of the host material, i.e., the GGI and ([Ag]+[Cu])/([Ga]+[In]) (or I/III) ratios, explaining the large discrepancies observed in the literature regarding the best cells' AAC ratio.

Ag-alloying has been tried in Cu(In,Ga)$S_2$ (sulfide) absorbers as well. Mori *et al.* showed improved absorber quality (of band gaps ranging from 1.54 eV to 1.66 eV) and larger short circuit current ($J_{SC}$), but also reported a lower open circuit voltage ($V_{OC}$) which ultimately led to a decreased cell efficiency. [19] Cheng *et al.* first prepared a (Ag,Cu)(In,Ga)$Se_2$ solar cell that showed improved performances and subsequently sulfurized the absorber, forming (Ag,Cu)(In,Ga)(S,Se)$_2$ ($E_g = 1.02$ eV) which further improved the performance. [20]

While previous research predominantly focused on selenide absorbers, the literature on sulfur-selenide absorbers alloyed with Ag remains to this day relatively sparse. This work aims to bridge this gap by investigating how Ag incorporation affects the elemental composition and performance of Cu(In,Ga)(S,Se)2 absorbers (CIGSSe). Different (low) amounts of Ag are introduced as precursor in the elemental stack of industrial CIGSSe. The bare absorbers are optically characterized, specifically by photoluminescence (PL) and cathodoluminescence (CL) spectroscopy, revealing an enhanced interdiffusion of Ga and In following Ag alloying. The absorbers' non-radiative losses are reduced, and the electrical performance of the resulting solar cells are measured.

## 2. Results and discussion

The effect of Ag-alloying on the film morphology and performances of CIGSSe absorbers is reported in the following sub-sections. The enhanced interdiffusion of In and Ga and the resulting lateral inhomogeneity is further discussed by means of photoluminescence, cathodoluminescence and Raman spectroscopy.

### 2.1. Effect on the crystal structure

Scanning electron microscope (SEM) imaging of the top surface and cross section of the samples is performed. **Figure 1** shows the images for the reference sample and the samples Ag-1 and Ag-3 (lowest and highest Ag content, see **Table 1** in the methods section). From the cross section images (Figure 1 bottom) an overall increase in the grain size is observed as more Ag is added, comparatively with the reference sample. With the highest Ag content, grains larger than 2 µm are obtained. This is expected as Ag has a lower melting point than Cu and thus growing (Ag,Cu)(In,Ga)(S,Se)$_2$ at similar temperature as the pure Cu-compound results in larger grains. [15] Moreover, in the reference sample, a layer of fine grains is present near the back contact. With increasing Ag content, the size of these fine grains increases until no strong contrast between them and the bulk ones is visible anymore. This is consistent with the observations reported in [14].

The top view SEM images (Figure 1 top) also reveal morphology modifications with added Ag. The reference sample exhibits a relatively smooth surface. With the lowest Ag content (Ag-1) crevices appear on the surface and further adding Ag (Ag-3) leads to an increased density of these crevices. Essig *et al.* reported similar feature in co-evaporated (Ag,Cu)(In,Ga)Se$_2$ absorbers, speculating that they are related to the agglomeration of Ag on the surface during deposition. [21] In the present case, the crevices seem to be non-detrimental from an optical characterization point of view, as will be discussed in the next section.

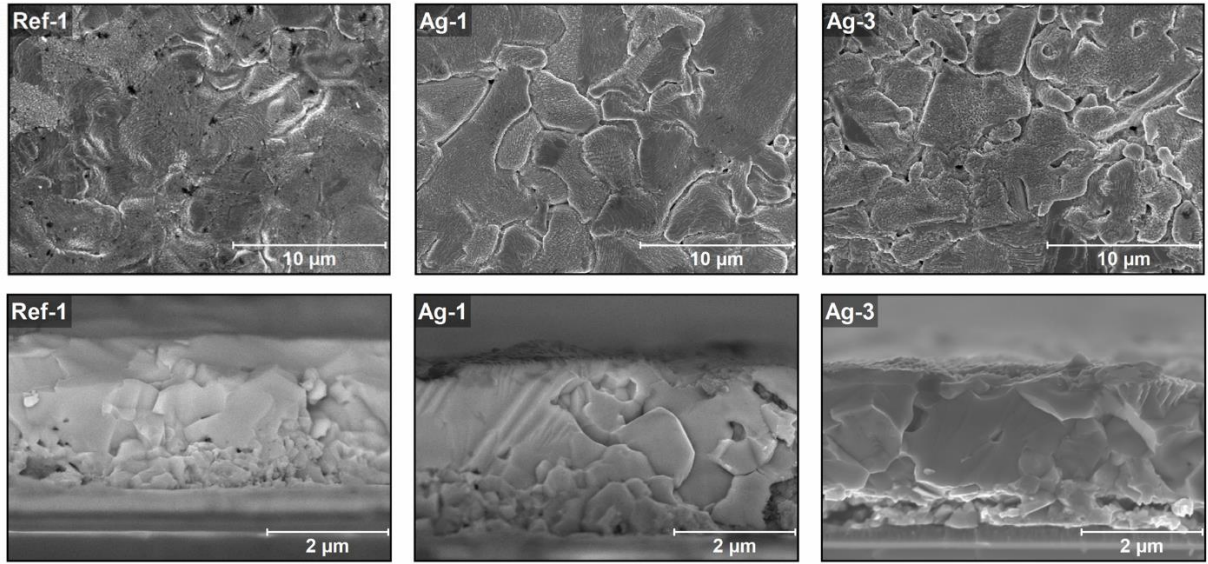

**Figure 1.** Scanning electron microscopy (SEM) images of top view (top) and cross section (bottom) of the reference absorber, containing no Ag, and two absorbers with low Ag content (Ag-1) and high Ag content (Ag-3). The sample Ag-3 is slightly detached from the substrate, most probably due to imperfect cleaving.

## 2.2. Composition investigation

The in-depth composition of the absorbers has been investigated by glow discharge optical emission spectroscopy (GDOES) and the GGI profiles of the three samples shown in Figure 1 are reported in **Figure 2**. The samples are grown with an intentional band gap gradient, leading to a profile like the reference one (green line): a low GGI near the front and an increasing GGI towards the back side. It has been demonstrated in a previous report (on similar CIGSSe absorbers without Ag) that in fact no smooth band gap increase generally exists. Instead, two phases of low GGI (low gap phase) and high GGI (high gap phase) form at the front and back side, respectively, and interlace in the bulk of the absorber. [22] This results in the average GGI profile depicted in Figure 2.

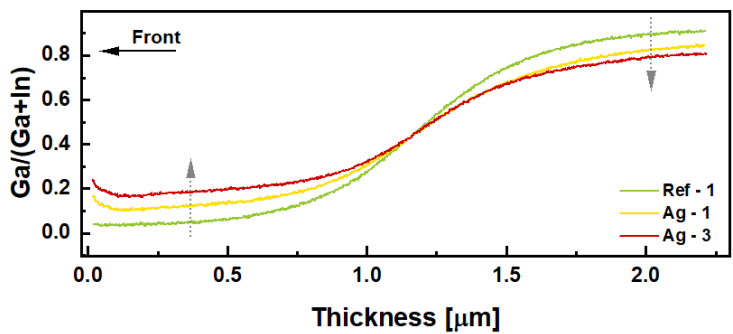

**Figure 2.** GGI profiles as measured by GDOES for the three absorbers shown in Figure 1. With increasing Ag content, the profiles flatten.

Upon addition of Ag, the GGI profiles flatten. The overall GGI at the notch position ($E_{g,min}$, from ~100 nm to 700 nm) increases from below 0.1 to 0.2 while the GGI at the back (over the last 400 nm) decreases from 0.9 to 0.8. Such flattening of the GGI profiles has been reported in low and high band gap Ag-alloyed chalcopyrites. [3,13] This suggests that Ag enhances the interdiffusion of Ga and In, as reported in [23]. Based on these profiles, a low gap phase of higher $E_g$ is expected towards the front surface, and a high gap phase of lower $E_g$ is expected near the back contact. This will be discussed in more details in sections 2.3 and 2.4. All the absorbers have been probed on four different spots and yield very similar profiles, attesting to the homogeneity of the samples. Exception is the sample Ag-3 for which the greatest discrepancies have been observed (**Figure S1.1** in the SI) hinting that higher Ag content may reduce the lateral homogeneity. Additionally, the S/(S+Se) profiles indicate a slight overall increase of the S content for all Ag-alloyed samples (**Figure S1.2** is SI).

The lateral inhomogeneity of the high-Ag sample is confirmed by Raman spectroscopy. **Figure 3** displays the spectra obtained from the front side of the samples Ag-1 (yellow) and Ag-3 (red) on different spots, compared to the reference sample without Ag (green). The latter shows two main peaks, associated to the A-like symmetry modes of the Se-Se and S-S vibrations, at 179 cm$^{-1}$ and 292 cm$^{-1}$, respectively. [24,25] The sample Ag-1 (lowest Ag content) yields the same two peaks, showing no significant difference to the reference sample. However, in the case of the sample Ag-3 (highest Ag content), these peaks are shifting in energy between different positions while scanning the surface, without showing any broadening. The Se-Se peak shifts from 179 cm$^{-1}$ to 184 cm$^{-1}$ together with the S-S peak from 292 cm$^{-1}$ to 298 cm$^{-1}$. Such a shift is usually associated with increased Ga content [26]. Thus, this observation could indicate the formation of (Ag,Cu)(In,Ga)(S,Se) phases of higher Ga content close to the surface in addition to the same low gap phase as the reference (producing the peaks at 179 cm$^{-1}$ and 292 cm$^{-1}$). The formation of phases of higher band gap explains the increase in the GGI ratio observed towards the front side.

It is usually observed that (Ag,Cu)(In,Ga)Se$_2$ solar cells have a low tolerance to (Ag,Cu)-poor off-stoichiometry composition [27] leading to the formation of ordered vacancy compounds (OVC). Keller *et al.* show that OVCs form for I/III ratios < 0.97. Above this value, other secondary phases start appearing as stoichiometry is approached. Both OVC and secondary phases are detrimental to the device performance. [7] In this study, no OVC signature in the Raman spectra is present in either sample.

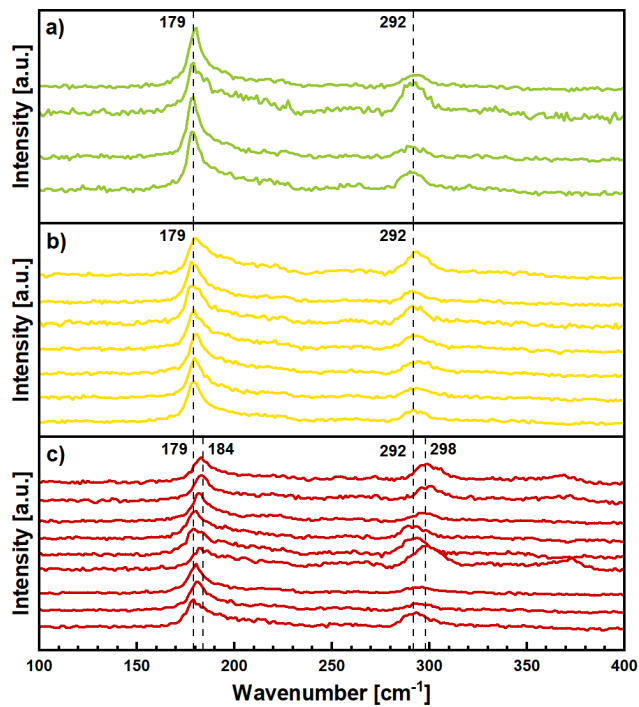

**Figure 3.** Raman spectroscopy performed on the reference sample a), the sample Ag-1 with the lowest Ag content b) and Ag-3 with the highest Ag content c). Each spectrum corresponds to a different spot on the sample. The spectra are normalized to the Se-Se peak and manually shifted for display purpose.

## 2.3. Photoluminescence spectroscopy

The bare absorbers are investigated by absolute photoluminescence from the front and back sides (see **Figure 4**). A soft KCN etching (5% concentration for 30 sec) is performed before measuring from the front side to remove oxides and refresh the surface after the absorbers have been exposed to air. The back side is accessed after performing a mechanical lift-off of the absorber from its substrate, similarly to in [22]. Photoluminescence spectra provide information about the recombination activity in the material. The reference sample in this study is a CIGSSe absorber with a band gap around 1.05 eV. Therefore, a dominant PL peak is expected at this specific energy and is attributed to band-to-band recombination (Figure 4a), green spectrum). As already deduced from the Raman analysis, the Ag-alloyed samples have the same low gap phase as the reference. The PL analysis confirms it since the same emission at 1.05 eV is observed for samples Ag-1 and Ag-3 from the front side. This is coherent with the theoretical band gap predictions for selenides computed in [3] for both low Ga and low Ag contents. However, with the highest Ag content (Figure 4a), red spectrum), a shoulder is seen in the spectrum at higher energies (~1.25 eV) from the front side and suggests the presence of an alternative recombination channel. The possibility of forming OVCs and/or secondary phases is discussed extensively for high band gap CIGSe in [27], reporting principally the formation of 1:3:5 OVCs ((Ag,Cu)(Ga,In)$_3$Se$_5$) at the front and back interfaces for I/III off-stoichiometry absorbers. Close to stoichiometry, the authors observed the formation of sparse Ag$_9$GaSe$_6$ grains, which are, however, mainly expected for above stoichiometry compounds. In the present case, only (A)CIGSSe phases and no OVC phases were detected from Raman spectroscopy, suggesting that the shoulder in the PL comes from an (A)CIGSSe phase of higher GGI than the low gap phase. The formation of this addtional phase reinforces the statement that Ga diffuses through the absorber with Ag alloying.

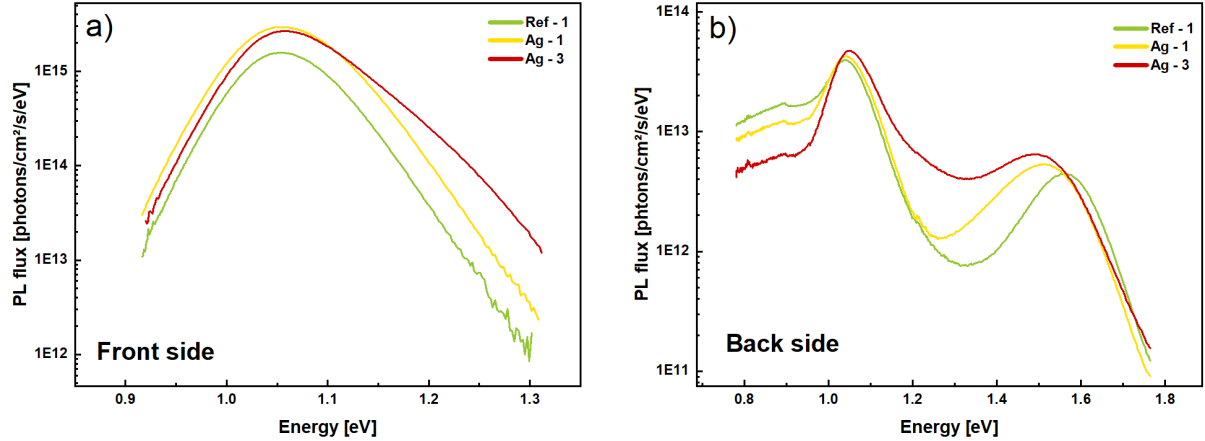

**Figure 4.** Absolute PL measured from the front a) and back b) sides of the reference sample and the samples Ag-1 and Ag-3. No high energy peak is observed from the front side and the part where $E > 1.3$ eV is not displayed. Note that the y-axes are in logarithmic scale.

Measurement from the back side reveals a second PL peak at 1.56 eV for the reference (Figure 4b), green spectrum). This second emission peak has been demonstrated to arise from the high gap phase forming near the back side of graded absorbers. [22] Although the three absorbers in Figure 4 were grown with the same amount of Ga and In precursors, the band gap of the high gap phase, dependent on the GGI at the back side, reduces with increasing Ag content (from 1.56 eV to 1.49 eV). Furthermore, the valley between the two peaks becomes shallower for higher Ag contents and the same shoulder at ~1.25 eV is detected for the sample Ag-3. This supports the earlier claim that the diffusion of Ga is enhanced by Ag alloying and additional evidence is provided by cathodoluminescence in section 2.4. The relation between the back side GGI and the PL maximum of the high gap phase follows well the trend previously observed in [22] (see **Figure S2** in SI). From the back side (Figure 4b)), a broad PL contribution is observed below 0.9 eV and attributed to a deep defect associated to the high gap phase forming near the back side of the absorbers. [22]

From absolute PL measurement on the front side, information about quasi-Fermi-level splitting (qFls), tail states, band gap fluctuations and optical diode factor (ODF) can be obtained and are discussed in the following. QFls gives an upper limit to the open circuit voltage ($V_{OC}$) and is therefore a very useful quantity to anticipate how good a finished cell can be. [28] Fitting the high energy part of the PL spectrum, sufficiently far above the band gap – where it is assumed that the absorptance is equal to 1 – one can get a measure of the qFls. [29] However, because of the shoulder in this range (see Figure 4a)), we opted for an evaluation of the qFls based on the PL quantum yield ($Y_{PL}$, referred as ERE in [29]) following Equation 1:

$$\text{qFls} = eV_{OC}^{SQ} - kT \cdot \ln(Y_{PL}) \tag{1}$$

with $V_{OC}^{SQ}$ the Shockley-Queisser $V_{OC}$ where the energy of the PL maximum is taken as the band gap and a perfect back reflector is considered. [30,31] $e$ is the elementary charge, $k$ is the Boltzmann constant, $T$ is the temperature and finally $Y_{PL}$ is the PL quantum yield defined as the ratio between the emitted photon flux and the absorbed one. Moreover, the quantity $kT \cdot \ln(Y_{PL})$ represents the non-radiative losses and, contrary to the qFls, does not depend on the band gap of the material, allowing for a comparison of absorbers with different band gaps. In addition to the three samples considered up to this point, four samples of intermediate Ag content but lower GGI in precursor stack are discussed (samples Ag-2.1 – Ag-2.4, see Table 1). Reducing the GGI has been demonstrated beneficial in other studies [22,32,33] and leads to further improvement in the present one. The GGI in the precursor stack of sample Ag-2.1 is lower than the one of the reference and is further reduced for the samples Ag-2.2, Ag-2.3 and Ag-2.4 (Table 1). Samples Ag-2.2 and Ag-2.3 are identical from a composition point of view, but the temperature of the rapid thermal process (RTP) of the second one is reduced. Finally, the sample Ag-2.4 is grown with a reduced Na content in the precursor stack.

**Figure 5** summarizes the results from the front side PL measurements. The highest non-radiative losses are recorded for the reference sample (188 meV). Alloying with Ag reduces these losses by 14 to 16 meV in the case of the samples Ag-1, Ag-3, Ag-2.1 and Ag-2.4 and a decrease by as much as 30 meV is achieved for the samples Ag-2.2 and Ag-2.3. Similar improvement would therefore be expected in the $V_{OC}$ of a cell made from these absorbers. As shown in **Figure S3.1** in the SI, an increase in the $V_{OC}$ is generally obtained (within error). However, the $V_{OC}$ improvement does not always correlate with the reduction of the non-radiative losses measure on the bare absorbers by PL. The $V_{OC}$ improvement for the sample Ag-2.2 is lesser than the expected one and no improvement for the sample Ag-2.3 is measured. This implies that additional losses may occur while completing the cell. Furthermore, the highest $V_{OC}$ is measured for the sample Ag-3 (relative improvement of 6.5% compared to the reference sample's $V_{OC}$) and even exceeds the corresponding qFls. Such behavior has been previously reported in [29]. Despite the higher $V_{OC}$, the cells did not show significant improvement in efficiency, mostly due to a reduction of the short circuit current density $J_{SC}$ (see Figure S3.1 in SI), leading to a lower product $J_{SC} \cdot V_{OC}$. The current losses may arise from the higher band gap observed at the front side, as measured from external quantum efficiency (EQE) (see **Figure S3.2** in SI) and further discussed in section 2.4. The best performing Ag-alloyed cell (based on absorber Ag-2.4) shows a relative improvement of 1.6% in average efficiency compared to the reference (slight improvement in $V_{OC}$ and $J_{SC}$ but decrease in FF). It can be expected that further optimization of the industrial processes for Ag containing absorbers will make full use of the reduced non-radiative recombination and result in improved efficiencies.

Interestingly, the change in the growth temperature of samples Ag-2.2 and Ag-2.3 did not affect the non-radiative losses. It is therefore possible to obtain similarly good absorbers at a lower process temperature, which might be specifically beneficial for industrial cells due to the cost reduction.

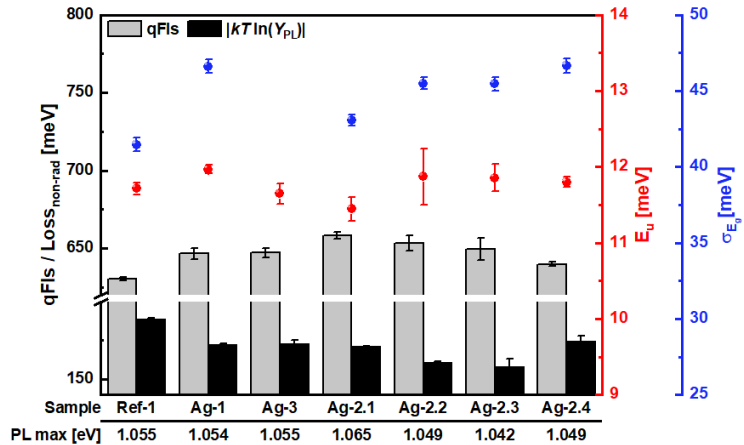

**Figure 5.** Summary of the PL results for seven samples. The samples Ag-2.1 – Ag-2.4 are samples with an Ag content between Ag-1 and Ag-3. The left axis is the qFls and the non-radiative losses defined by $|kT \cdot \ln(Y_{PL})|$. The first right axis (red) is the Urbach energy and the second one (blue) is the standard deviation from the mean band gap as defined in Equation 2. The energy of the PL peak maximum (used to calculate the qFls) is given for each sample below the bottom axis.

The low energy part of the PL spectrum, far below the band gap, can be used to measure the tail states. [34] Tail states arise from structural disorder, band gap fluctuations or electrostatic fluctuations and represent a density of states extending into the forbidden gap. They have a direct detrimental impact on $V_{OC}$. [35–37] The density of tail states decreases exponentially from the band edge and can be described by the Urbach energy $E_U$. A reduction of the Urbach energy upon incorporation of Ag is observed in wide band gap CIGSe [2,5], but the opposite is reported for low band gap CIGSe. [13] In the present study, no significant change is observed (see Figure 5). The measured $E_U$ range from 11.4 meV to 12.0 meV and $E_U^{\text{Ref}} = 11.7$ meV. This suggests that the tail states are not reduced by the Ag incorporation. Another parameter linked to the disorder is the sharpness of the absorption edge. In fact, the PL emission depends on the distribution of band gaps in the material. For samples without a band gap gradient, a sharp absorption edge is expected, but with a gradient or if the alloy disorder is increased, the absorption edge becomes less sharp. [38–40] Assuming a Gaussian distribution of the energy of step like band gaps, one can fit the derivative of the absorptance and extract the standard deviation from

$$\frac{dA(E)}{dE} = C_0 + \frac{C_1}{\sigma_{E_g} \cdot \sqrt{2\pi}} \exp\left(-\frac{(E - E_g)^2}{2\sigma_{E_g}}\right) \quad (2)$$

where $A(E)$ is the absorptance (**Figure 6**), $E_g$ corresponds to the mean band gap and $\sigma_{E_g}$ is the standard deviation. $C_0$ and $C_1$ are fitting constants. The absorptance is obtained from the PL measurement by solving the generalized Planck's law for $A(E)$. [29,34]

The standard deviation for each of the samples is reported in Figure 5. The smallest one is obtained for the reference (41 meV) and increases up to 47 meV for the samples Ag-1 and Ag-2.4. Due to the shoulder at 1.25 eV in the PL spectrum, the absorptance of the sample Ag-3 – which showed the largest inhomogeneity in the GGI profile and in the Raman spectra – could not be extracted. Such low spreading in the standard deviation (6 meV) is not significant and does not support firm conclusions. The changes observed in the band gap gradient are too small to have an effect on the broadening. Other effects seem to be more important.

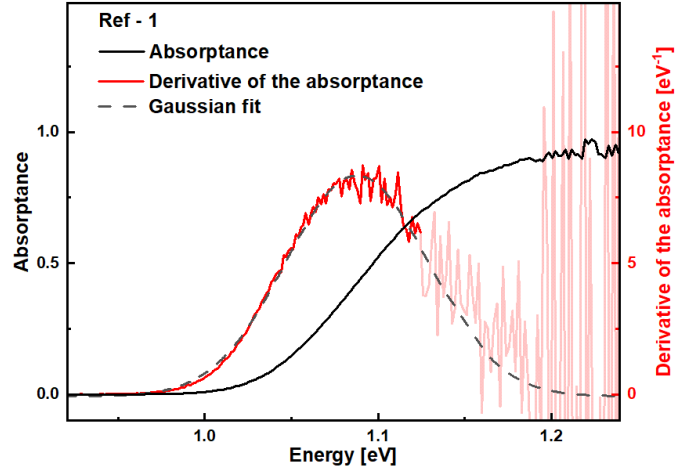

**Figure 6.** Absorptance obtained from a PL measurement on the reference sample (black curve) and its derivative (red curve). The derivative is fitted with Equation 2 (dashed line) over the highlighted range (0.90 – 1.14 eV). The high energy part of the derivative is disregarded because of the high noise.

Finally, a decisive parameter for solar cell efficiency is given by the fill factor (FF). The record CIGS solar cell reaches a FF just above 80%. [1] The diode factor is originally defined in the diode equation $j = j_0 \left( \exp\left(\frac{eV}{A_{el}kT}\right) - 1 \right)$ where $j$ is the current density, $j_0$ is the saturation current density, $V$ is the voltage, $e$ is the elementary charge, $k$ is the Boltzmann constant, $T$ is the temperature and $A_{el}$ is the electrical diode factor (e.g. in [41]). The latter has a direct influence on the fill factor of the solar cell and takes generally a value between 1 and 2, depending on the recombination mechanism. For a good FF the diode factor has to be as low as possible. A diode factor larger than 1 is usually considered indicative of recombination in the space charge region (SCR). Empirically, it has been observed that the PL flux $\phi_{PL}$ depends on the generation flux $G$ and follows the power law $\phi_{PL} \propto G^{A_{opt}}$. Trupke et al. [42] and later Babbe et al. [43] demonstrate that the exponent $A_{opt}$ equals the electrical diode factor $A_{el}$ under two conditions. First, a good collection of the photogenerated charges must be assumed and second, the quasi Fermi level splitting measured optically must equal the $V_{OC}$ of the cell. In all other cases, the electrical diode factor will be larger than the optical one. Measuring excitation-dependent photoluminescence thus leads to a determination of the optical diode factor (ODF). From the slope of a linear fit in a logarithmic scale, one can read a value for the ODF (see **Figure S4** in SI). It is found that upon Ag incorporation, the ODF of the bare absorbers remains unchanged and equals the one of the reference sample, ODF = 1.3. It is therefore reasonable to expect similar FF for cells prepared from these absorbers, if the subsequent deposited layers (buffer, window layers) are not affected by the different Ag contents themselves. The amount of Ag used in the present absorbers is small enough that no significant change of the band alignment is expected. [3]

In the case of a bare absorber – without buffer – there exists no SCR. Weiss *et al.* [44,45] show that an ODF larger than 1 is explained by an additional shift of the majority carriers quasi Fermi level, even in low injection condition. Upon illumination, metastable defects, like the double vacancy complex $V_{Se} - V_{Cu}$ in CIGS [46], can change from a donor state to an acceptor state, thus increasing the net acceptor density and in turn shifting the hole quasi Fermi level closer to the valence band.

### 2.4. Cathodoluminescence spectroscopy

In the previous section, interdiffusion of Ga and In, as well as lateral inhomogeneity arising upon Ag alloying have been discussed. This section focuses on CL performed on the cross section of the absorber Ag-3 and provides further evidence of the above-mentioned observations. A CIGSSe sample similar to the reference from this study has already been investigated in another report [22] and some of the major results are reproduced in **Figure 7**. The CL results for the sample Ag-1 can be found in **Figure S5.1** in the SI. A spectrum is measured for each pixel in the CL measurement, creating a hyperspectral dataset. The corresponding panchromatic emission intensity is depicted in Figure 7b) and the emission energy of the intensity maximum is reported in c). It is important at this point to mention that the shown CL measurements were performed using an uncorrected silicon CCD detector. This detector is characterized by a significant reduction in detection efficiency towards the lower measured energies. Measurement performed by an InGaAs diode array, with enhanced sensitivity at the low energy scale can be found in the SI (**Figure S5.2**) and shows a brighter front side, as expected. This explains the lower intensity of the signal near the front surface (Figure 7b)), as well as the discrepancy between CL and PL about the energy of the luminescence maximum of the low gap phase (1.05 eV from PL and 1.15 eV from CL) since PL utilizes an InGaAs detector in this range. The high energy emission agree better as both methods use a silicon detector in this range.

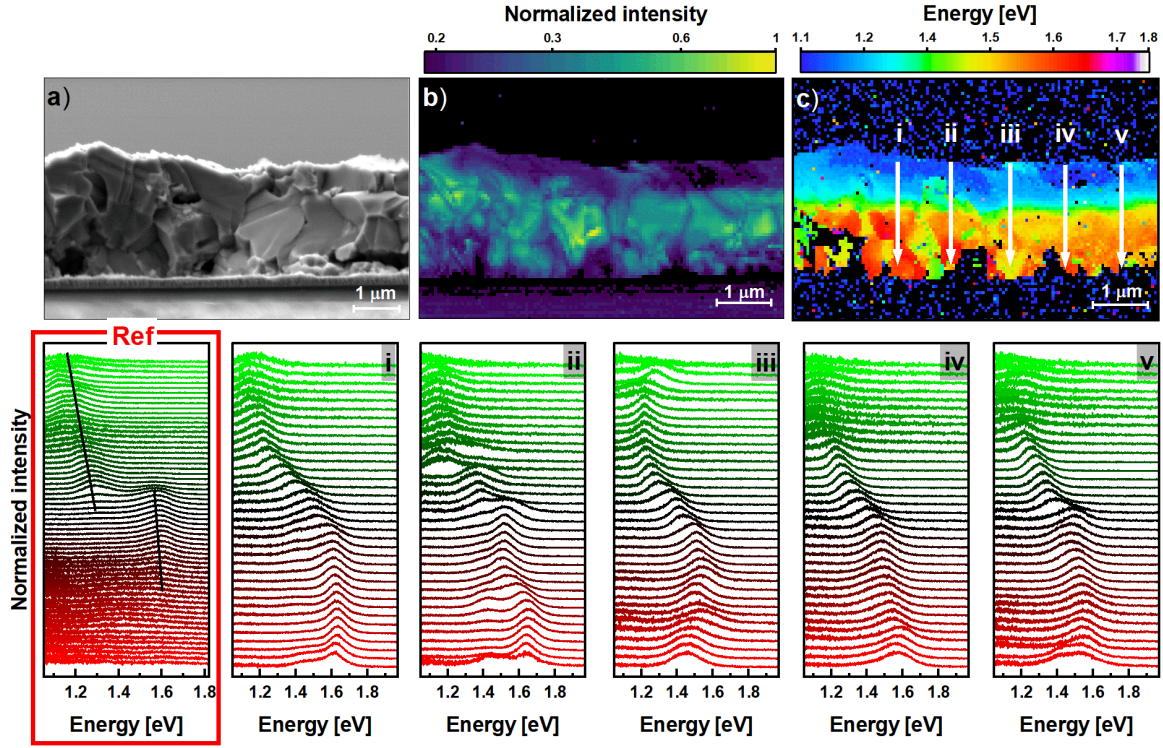

**Figure 7.** Cathodoluminescence spectroscopy on the cross section of the sample Ag-3. The top row shows the SE image of the region of interest a), the normalized intensity map of the panchromatic luminescence in logarithmic scale b) and the colormap of the energy of the maximum of the CL emission c). The bottom row shows the individual spectra along the lines **i-v** drawn in c). The leftmost graph in the bottom row is reproduced from [22] and corresponds to typical emission observed in the reference sample. The black lines are a guide to the eye. All the spectra are normalized and shifted for clarity.

Although each pixel in the colormaps is of one single color, the transition from one $E_g$ to the other is often not sharp and two peaks are visible, but only the energy of the dominant one is displayed. A sudden jump in emission energy from the low band gap phase to the high band gap phase has been the most frequent observation in [22] and an example is given in Figure 7 (red-framed graph). From the front surface towards the back contact, $E_g$ increases only slightly until a certain depth in the bulk where a jump to a higher $E_g$ occurs. It is visible from the two peaks present in the spectra at this depth and highlighted by the dark lines. This led to the conclusion that in graded absorbers, there is no smooth band gap gradient, but instead that a low gap phase forms towards the front, a high gap phase forms towards the back, and the two phases interlace in the bulk of the material.

Similar behavior is observed for the line profile **ii** in Figure 7, although it is not the most common situation. Indeed, the line profiles **i** and **iii-v** reveal a smoother gradient from the front side to the back side, suggesting that Ag increases the gradient behavior, likely by Ga interdiffusion. However, a variation of band gaps close to the back side is also observed (spectra with two clearly identifiable peaks, e.g. line profiles **ii** and **v**). This inhomogeneity is translated in the energy map (Figure 7c)) by the abrupt change in color. The lateral inhomogeneity is highlighted by the intermittence of green/yellow and red close to the back side. Additionally, the first few hundreds of nanometers from the front surface show a mix of dark and light blues, indicating the variation of band gaps as observed from Raman and PL. In particular, patches of higher band gap (cyan-green) are present near the front surface (also visible in the top spectra from the line profile **iii** in Figure 7). For the sample Ag-1 (lowest Ag content, Figure S5.1 in SI), larger lateral inhomogeneity is observed, and more energy jumps appear in the line profiles.

Taken together, the cathodoluminescence results support the increased diffusion of Ga from the back side towards the front as suggested by the variation of band gaps near the front and back surfaces. This in turns, provides the explanation for the flattening of the GGI profiles observed by GDOES. With the formation of higher (lower) band gap phases near the front (back) surface, the average GGI in this region is expected to increase (decrease). Moreover, with increasing Ag content, a smoother band gap gradient is achieved, despite a certain lateral inhomogeneity.

## 3. Conclusion

Ag-alloyed industrial chalcopyrites $(Ag,Cu)(In,Ga)(S,Se)_2$ absorbers have been grown with different (low) Ag contents. It has been found that the crystal morphology improves with Ag, in particular showing an increased grain size. Upon Ag-alloying, a stronger interdiffusion of Ga and In is observed, resulting in a redshift of the high energy peak measured by photoluminescence. This correlates with a decrease of the average back side GGI, measured by GDOES. As the back side GGI decreases, the front side GGI increases, leading to a flattening of the GGI profile. The increase of the GGI towards the front side is attributed to the formation of a (A)CIGSSe phase of higher Ga content, which is confirmed by Raman spectroscopy. Raman spectroscopy indicates more lateral inhomogeneity with increasing Ag. Cathodoluminescence spectroscopy on the cross section of the samples supports the lateral inhomogeneity. However, a smoother band gap gradient is obtained after Ag alloying compared to the reference. Finally, a reduction of the non-radiative losses of up to 30 meV is measured by PL on the bare absorbers, which translated into a $V_{OC}$ increase of the corresponding solar cells.

## 4. Methods

*Sample preparation*: The absorbers investigated in this study are produced in the pilot line of AVANCIS, Germany, according to the SEL-RTP (stack elemental layer followed by rapid thermal processing). [47,48] They are grown based on a double barrier back-electrode for the control of alkali-diffusion and Mo selenization. The elemental precursors (Ag)-Cu-In-Ga:Na are sputtered and Se is thermally evaporated. The stack undergoes a rapid thermal process in S atmosphere, completing the formation of the $(Ag,Cu)(In,Ga)(S,Se)_2$ absorber. The absorbers are grown with same elemental stack precursors but differ in the amount of Ag introduced. The reference sample (AAC = 0) and samples of three different Ag contents are grown with similar CGI. The Ag content is kept low for every variation and does not exceed 10%, i.e., AAC < 0.1. In the following, the names Ag-1 and Ag-3 refer to the lowest and highest Ag content, respectively (see Table 1).

**Table 1.** List of the samples from this study.

| Sample | GGI | AAC | Description |
|---|---|---|---|
| Ref-1 | **GGI-3** | **No Ag** | Reference |
| Ag-1 | | **AAC-1** | Lowest AAC |
| Ag-3 | | **AAC-3** > AAC-1 | Highest AAC |
| Ag-2.1 | **GGI-2** < GGI-3 | AAC-1 < **AAC-2** < AAC-3 | Intermediate AAC, Intermediate GGI |
| Ag-2.2 | **GGI-1** < GGI-2 | | Intermediate AAC, Lowest GGI |
| Ag-2.3 | | | Intermediate AAC, Lowest GGI, Lower process temperature |
| Ag-2.4 | | | Intermediate AAC, Lowest GGI, Lower Na in elemental stack |

*Photoluminescence spectroscopy*: The absorbers are measured by absolute photoluminescence spectroscopy. The experimental setup is first calibrated and the resulting spectra are corrected spectrally and in intensity, as described in [22]. Two detectors are utilized to cover a larger spectral window, ranging from 400 nm to 900 nm using a CCD Si camera (Andor iDus DV420A-OE) and from 900 nm to 1600 nm using an InGaAs detector (Andor iDus DU490A-1.7). The broadest measured spectrum extends however only from 700 nm to 1600 nm (~0.78 eV to 1.77 eV) as a long-pass filter with cut-off wavelength 700 nm is introduced in the setup to protect the detectors from the 660 nm wavelength diode laser used as excitation source. If not otherwise stated, the absolute PL measurements are performed under a photon flux of $2.85 \times 10^{17}$ cm$^{-2}$s$^{-1}$, corresponding to a one-sun illumination for an absorber of band gap of 1.05 eV based on the AM1.5g solar spectrum. The numerical results (and error bars) obtained by PL are averaged over three different spots on each absorber. The InGaAs detector used has two diodes of slightly different sensitivity. Therefore, in order to reduce the noise in the PL spectra and corresponding absorptance, only the least noisy of the two is considered.

*Cathodoluminescence spectroscopy*: CL hyperspectral mapping is performed at 300 K on a cleaved cross-section in an Attolight Allalin 4027 Chronos SEM-CL system. CL measurements are taken using an iHR320 spectrometer with a grating density of 150 lines per mm blazed at 500 nm. The microscope is operated at an electron beam current of 2.5 nA and an electron landing energy 5 keV. The CL hyperspectral maps are then analyzed using LumiSpy. [49]

*Raman spectroscopy*: Raman spectroscopy was acquired with a Renishaw inVia micro-Raman spectrometer equipped with a 532 nm excitation laser source, a 100× objective lens (with a numerical aperture of 0.85) and a 2400 lines/mm grating. Several spots are measured in each case, spaced by a few microns.

*Scanning electron microscope*: The morphology and thickness of the bare absorbers were assessed with a Hitachi SU-70 field-emission SEM. Top-view and cross-sectional SEM images were taken with a voltage of 7 kV.

**Supporting information**
Supporting Information is available from the Wiley Online Library or from the author. The supplementary material provides complementary data as indicated in the text: complementary GGI and S/(S+Se) profiles from GDOES; relation between the back side GGI and the PL from the back side; relative IV parameters and EQE of the finished cells; optical diode factor measurement; CL measurement on the sample Ag-1 similar to the one of sample Ag-3 shown in Figure 7 and CL measurement with an InGaAs detector of the sample Ag-3.

## Acknowledgments

This work has been supported by Avancis, Germany, in the framework of the POLCA project, and by EPSRC in the framework of the REACH project (EP/V029231/1), which are gratefully acknowledged. The authors thank Dr. Mohit Sood for his suggestions and discussion.

## Conflict of interest

The authors declare no conflict of interest.

## Author contributions

**Aubin JC. M. Prot:** Conceptualization (equal); Investigation (lead); Formal analysis (lead); Visualization (lead); Writing/Original Draft Preparation (lead); Writing/Review & Editing (equal). **Michele Melchiorre:** Investigation (supporting); Formal analysis (supporting). **Tilly Schaaf:** Investigation (supporting). **Ricardo G. Poeira:** Investigation (supporting); Writing/Review & Editing (equal). **Hossam Elanzeery:** Formal analysis (equal); Investigation (equal); Resources (equal); Writing/Review & Editing (equal). **Alberto Lomuscio**: Formal analysis (equal); Investigation (equal); Resources (equal); Writing/Review & Editing (equal). **Souhaib Oueslati:** Formal analysis (equal); Investigation (equal); Resources (equal); **Anastasia Zelenina:** Formal analysis (equal); Investigation (equal); Resources (equal); Writing/Review & Editing (equal). **Thomas Dalibor**: Formal analysis (equal); Resources (equal); Writing/Review & Editing (equal); Funding Acquisition (lead). **Gunnar Kusch**: Formal analysis (equal); Investigation (equal); Writing/Review & Editing (equal). **Yucheng Hu**: Formal analysis (equal); Investigation (equal). **Rachel A. Oliver**: Conceptualization (lead); Formal analysis (equal); Writing/Review & Editing (equal); Funding Acquisition (lead). **Susanne Siebentritt**: Conceptualization (lead); Formal analysis (equal); Funding Acquisition (equal); Supervision (lead); Writing/Review & Editing (equal).

## Data availability

The data supporting the findings of this study is openly available in Zenodo at 10.5281/zenodo.10686926.